\documentclass[twocolumn,amsmath,amssymb,floatfix]{revtex4}

\usepackage{graphicx}
\usepackage{dcolumn}
\usepackage{bm}

\graphicspath{{Figures/}}
\setkeys{Gin}{width=\linewidth}

\begin{document}

\title{Transport through graphene double dots}

\author{F. Molitor, S. Dr\"oscher, J. G\"uttinger, A. Jacobsen, C. Stampfer, T. Ihn and K. Ensslin}

\affiliation{Solid State Physics Laboratory - ETH Zurich, Switzerland}

\begin{abstract}

We present Coulomb blockade measurements in a graphene double dot system. The coupling of the dots to the leads and between the dots can be tuned by graphene in-plane gates. The coupling is a non-monotonic function of the gate voltage. Using a purely capacitive model, we extract all relevant energy scales of the double dot system.

\end{abstract}

\maketitle


The control of individual electrons and spins \cite{Elzerman2004,Fujisawa2002,Ono2002} has been achieved in semiconductor quantum dots. Graphene as a material systems lends itself for small and well controlled quantum systems \cite{Stampfer2008,Ponomarenko2008} with the additional possible benefit of increased spin coherence times \cite{Trauzettel2007}. Controlled coupling of quantum dots is a prerequisite for the envisioned implementation of spin qubits \cite{Loss1998} in such systems.

Here we demonstrate Coulomb blockade of a graphene double dot system. With a model of purely capacitively coupled dots \cite{VanderWiel2008} we extract all the characteristic energy scales of the double dot system. The tunnel barriers between the dots and the source or drain contact, as well as the tunnel barrier separating the two dots can be tuned by graphene in-plane gates. We show that the tunnel coupling is a non-monotonic function of the plunger gate voltage as expected from transport experiments on gated graphene nanoribbons \cite{Stampfer2009,Molitor2009,Todd2008,Lan2008}.


\begin{figure}
  \begin{center}
    \includegraphics{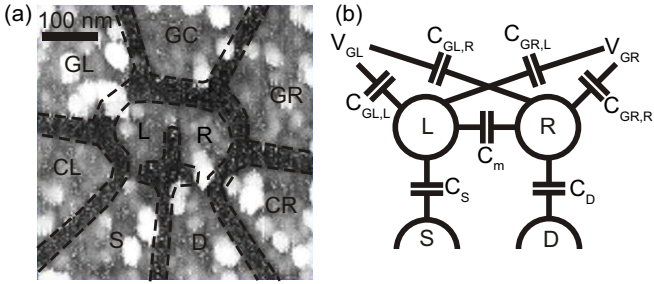}
    \caption{(a) Scanning force micrograph (SFM) of the double dot structure studied in this work. The outline of the graphene regions is highlighted by the dashed lines. The dots, labelled by L and R, have a diameter of 90 nm and the constriction between them has a lithographic width of 30 nm. The dots are connected by 20 nm wide constrictions to source and drain contacts. The gates, labelled by CL, GL, GC, GR and CR, are located 40 nm from the structure. (b) Model for the analysis of the double dot system. $C_{\mathrm{m}}$ is the mutual capacitive coupling between the dots. $C_{\mathrm{GL,L}}$ and $C_{\mathrm{GR,R}}$ are the capacitances coupling the electrochemical potential in the left and right dot to the voltages $V_{\mathrm{GL}}$ and $V_{\mathrm{GR}}$ applied to their respective gates labelled by GL and GR in Fig. (a). $C_{\mathrm{GR,L}}$ and $C_{\mathrm{GL,R}}$ are the capacitances between the right gate and the left dot and vice-versa. The capacitances $C_{\mathrm{S}}$ and $C_{\mathrm{D}}$ describe the electrostatic coupling of the source and drain contacts to the double dot.}
    \label{fig1}
  \end{center}
\end{figure}

The graphene flakes are produced by mechanical cleaving of natural graphite flakes and deposition on a highly doped silicon substrate covered by 285~nm of silicon dioxide \cite{novoselov_science2004}. Thin flakes are found by optical microscopy. Raman spectroscopy is used to select single-layer flakes \cite{ferrari_prl2006,graf_nanolett2007}. The contacts are defined by electron beam lithography, followed by the evaporation of Cr/Au (2~nm/40~nm). Finally the graphene flake is patterned by defining the structure with electron beam lithography using a 45~nm thick PMMA resist layer, followed by reactive ion etching based on Ar and $\mathrm{O}_{2}$ (2:1) \cite{guttinger2009}. 

Fig. \ref{fig1}(a) shows a scanning force micrograph of the double dot structure studied in this work. The structure consists of two central graphene islands forming the dots labeled L and R in Fig. \ref{fig1} (a). They are mutually connected by a 30~nm wide constriction. Each dot has a diameter of about 90 nm. The dots are connected by 20~nm wide constrictions to source and drain contacts. In addition to the doped substrate which acts as a global back gate, there are five graphene in-plane gates allowing to fine-tune the structure. The gates GL and GR can be used to change the number of carriers in the dots, while the gates CL, CR and GC are used to tune the transmission of the constrictions and the coupling between the dots. For all gates, no leakage currents can be detected for applied voltages up to $\pm$10~V.

All the measurements presented in this paper are recorded at a temperature of 1.4~K, with 0~V applied between the back gate and the graphene double quantum dot circuit.


\begin{figure}
  \begin{center}
    \includegraphics{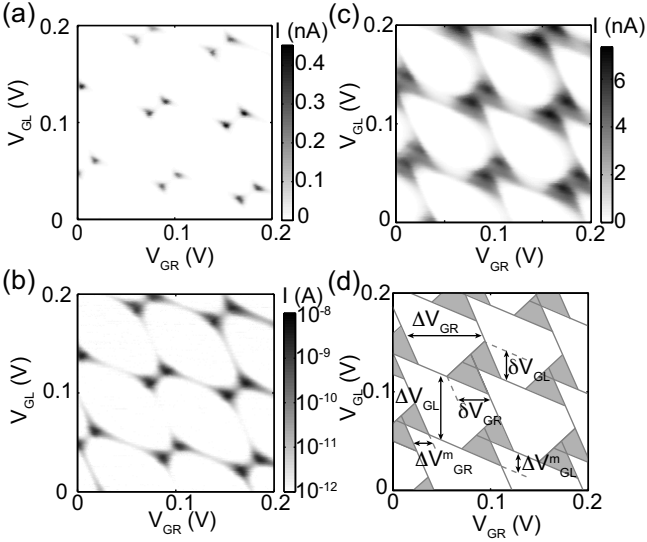}
    \caption{Current through the double dot as a function of the voltages $V_{\mathrm{GR}}$ and $V_{\mathrm{GL}}$ applied to the gates GR and GL for $V_{\mathrm{BG}}=0$~V, $V_{\mathrm{CL}}=-4.65$~V, $V_{\mathrm{GC}}=-1.2$~V and $V_{\mathrm{CR}}=0$~V. (a),(b): Representation in linear (a) and logarithmic (b) scale for a bias voltage $V_{\mathrm{bias}}=500~\mu$V applied between source and drain contacts. (c) Representation in linear scale for a bias voltage $V_{\mathrm{bias}}=-4$~mV. (d) Illustration of the measurement displayed in (c), with annotation of the quantities used to deduce the energy scales of the system.}
    \label{fig2}
  \end{center}
\end{figure}

Fig. \ref{fig2} (a) shows a measurement of the current through the double dot as a function of the voltages applied to gates GR and GL for an applied bias voltage $V_{\mathrm{bias}}=500~\mu$V. The honeycomb pattern characteristic for the charge stability diagram of a double dot \cite{VanderWiel2008} can be observed. Elastic transport through the double dot is only possible in the case where the electrochemical potentials in both dots are aligned mutually and with the Fermi energy in the leads. This is the case at the so-called triple points in the corners of the hexagons of constant charge configuration. A plot of the same measurement displaying the current on a  logarithmic scale [Fig. \ref{fig2} (b)] makes the connecting lines between the triple points visible. Along these lines, only one of the dot levels is aligned with the Fermi energy in the leads, leading to current by co-tunneling processes. The current through the edges of the hexagons can be suppressed by changing the voltages applied to gates CL and CR in such a way that the barriers are less transparent.

When applying a higher bias voltage $V_{\mathrm{bias}}=-4$~mV between source and drain contacts, the triple points evolve into triangular-shaped regions of increased current \cite{VanderWiel2008} [Fig. \ref{fig2} (c)]. We use the model of purely capacitively coupled dots, presented in Fig. \ref{fig1} (b), to estimate the energy scales of the system, assuming that the applied source-drain voltage drops entirely over the double dot system. Fig. \ref{fig2} (d) shows a schematic of the measurement from Fig. \ref{fig2} (c), with indication of the different parameters used in the following to estimate the energy scales of the system. The extension of the triangular-shaped regions allow the determination of the conversion factors between gate voltage and energy. The lever arm between the left gate GL and the left dot is $\alpha_{\mathrm{GL,L}}=V_{bias}/\delta V_{\mathrm{GL}}=0.13$ and between the right gate GR and the right dot $\alpha_{\mathrm{GR,R}}=V_{bias}/\delta V_{\mathrm{GR}}=0.12$. The lever arms between the left gate and the right dot and vice-versa are determined from the slope of the co-tunneling lines delimiting the hexagons: $\alpha_{\mathrm{GL,R}}=0.06$ and $\alpha_{\mathrm{GR,L}}=0.05$. The dimensions of the honeycomb cells $\Delta V_{\mathrm{GL}}$ and $\Delta V_{\mathrm{GR}}$ give the capacitances between the gate GL and the left dot $C_{\mathrm{GL}}=e/\Delta V_{\mathrm{GL}}=2.4$~aF, and between GR and the right dot $C_{\mathrm{GR}}=e/\Delta V_{\mathrm{GR}}=2.0$~aF. The corresponding total capacitances of the dots are $C_{\mathrm{L}}=C_{\mathrm{GL}}/\alpha_{L}=18.0$~aF and $C_{\mathrm{R}}=C_{\mathrm{GR}}/\alpha_{R}=17.5$~aF. This corresponds to single-dot charging energies $E_{\mathrm{C}}^{L}=\alpha_{\mathrm{GL,L}}\cdot\Delta V_{\mathrm{GL}}=8.9$~meV and $E_{\mathrm{C}}^{R}=\alpha_{\mathrm{GR,R}}\cdot\Delta V_{\mathrm{GR}}=9.2$~meV, which is comparable to the values found for graphene single dots of similar size \cite{schnez_2009}. The coupling energy between both dots can be extracted from the splitting of the triple points: $E_{\mathrm{C}}^{m}=\alpha_{\mathrm{GL,L}}\cdot \Delta V^{\mathrm{m}}_{\mathrm{GL}}=\alpha_{\mathrm{GR,R}}\cdot \Delta V^{\mathrm{m}}_{\mathrm{GR}}=2.5$~meV. For this temperature and gate voltage range the tunnel coupling energy between the two dots is below the experimental resolution.


\begin{figure}
  \begin{center}
    \includegraphics{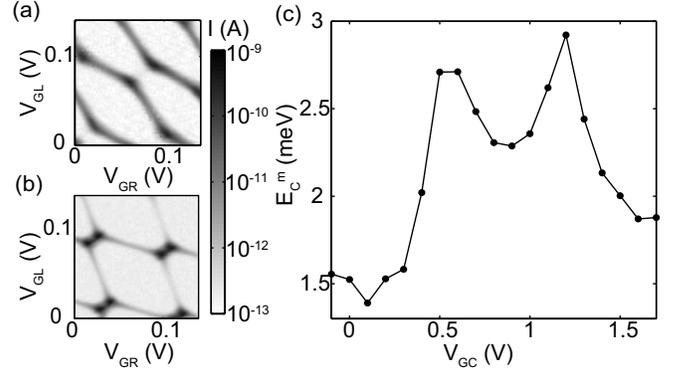}
    \caption{(a),(b): Examples for two extreme cases of strong (a) and weak (b) mutual coupling between the dots. The voltages applied to the gates are $V_{\mathrm{BG}}=0$~V, $V_{\mathrm{CL}}=-4.65$~V and $V_{\mathrm{CR}}=0$~V , $V_{\mathrm{GC}}=-1.9$~V (a) and $V_{\mathrm{GC}}=0$~V (b). (c) Coupling energy as a function of the voltage $V_{\mathrm{GC}}$ applied to the gate GC. The voltages applied to the other gates are $V_{\mathrm{BG}}=0$~V, $V_{\mathrm{CL}}=-4.65$~V and $V_{\mathrm{CR}}=1$~V. The ranges of $V_{\mathrm{GL}}$ and $V_{\mathrm{GR}}$ are adjusted for each value of $V_{\mathrm{GC}}$ using appropriate values based on the measured lever arms in order to always stay always at the same triple point.}
    \label{fig3}
  \end{center}
\end{figure}

By changing the voltage applied to the central plunger gate labelled GC in Fig. \ref{fig1} (a), we are able to change the coupling between both dots. Fig. \ref{fig3} (a,b) show an example of the charge stability diagram for two extreme cases of strong (a) and weak (b) coupling between the dots. The only difference between both measurements is the voltage applied to the gate GC [(a): $V_{\mathrm{GC}}=-1.9$~V, (b): $V_{\mathrm{GC}}=0$~V]. The coupling energy between the dots changes by more than a factor of two: (a): $E_{\mathrm{C}}^{m}=4.2$~meV (b): $E_{\mathrm{C}}^{m}=1.7$~meV. The single-dot charging energies do not change significantly compared to the case shown in Fig. \ref{fig2}, assuming the lever arms are still the same.

Fig. \ref{fig3} (c) presents a closer analysis of how the coupling energy changes with $V_{\mathrm{GC}}$. It displays the coupling energy as a function of the voltage $V_{\mathrm{GC}}$ for the same triple point, followed through the $V_{\mathrm{GR}}$ - $V_{\mathrm{GL}}$ parameter space as $V_{\mathrm{GC}}$ is changed. The lever arms are assumed to be the same as in Fig. \ref{fig2}. The strength of the inter-dot coupling shows a non-monotonic behavior as a function of applied $V_{\mathrm{GC}}$: the coupling energy starts at quite small values, increases by a factor of two for more positive $V_{\mathrm{GC}}$, before decreasing again. This is in agreement with recent experiments on graphene nanoribbons \cite{Stampfer2009,Molitor2009,Todd2008,Lan2008}, showing a strongly non-monotonic dependence of the current on gate voltage, with many sharp resonances. The conductance has been shown to vary over orders of magnitude as a function of in-plane gate voltage \cite{Stampfer2009,Molitor2009}. Here, we rather probe the electrostatic landscape between the two dots and the corresponding change in coupling is only a factor of two for a given resonance. It remains to be seen how the tunneling transmission of the constriction can be linked to the electrostatic coupling it provides between two quantum systems.

In conclusion, we have demonstrated Coulomb blockade in a graphene double dot system. The coupling between both dots, as well as the transmission of the constrictions connecting the dots to the leads, can be tuned by graphene in-plane gates. We have shown that the coupling between the dots is a non-monotonic function of the applied gate voltage. Finally, a model of purely capacitively coupled dots allowed to extract the relevant energy scales of the system. The presented results may be seen as a promising development towards the realization of spin qubits in graphene.

We thank C. Barengo and M. Csontos for help with the setup, T. Choi and B. K\"ung for helpful discussions and the Swiss National Science Foundation (SNF) and NCCR Nanoscience for financial support.



\end{document}